\documentclass[a4paper,11pt]{article}

\usepackage{bm}
\usepackage{latexsym}
\usepackage{amssymb}
\usepackage{amsmath}
\usepackage{mathtools}
\usepackage{hyperref}
\hypersetup{colorlinks=true, allcolors=blue}
\usepackage{graphicx}

\usepackage{mathbbol}
\usepackage{braket}
\usepackage{empheq}
\usepackage{cite}
\usepackage[title]{appendix}
\usepackage{here}
\usepackage[normalem]{ulem}
\usepackage{pifont}

\makeatletter
\renewcommand\subparagraph{
    \@startsection {subparagraph}{5}{\z@ }{3.25ex \@plus 1ex
    \@minus .2ex}{-1em}{\normalfont \normalsize \bfseries }}
\makeatother

\numberwithin{equation}{section}

\begin{document}
\pagestyle{empty}

\vspace{-4cm}
\begin{center}
    \hfill KEK-TH-2762, YITP-25-195 \\
\end{center}

\vspace{2cm}

\begin{center}

{\bf\LARGE  
Probing high-frequency gravitational waves with entangled vibrational
qubits in linear Paul traps\\
}

\vspace*{1.5cm}
{\large 
Ryoto Takai
} \\
\vspace*{0.5cm}

{\it 
KEK Theory Center, Tsukuba 305-0801, Japan\\
The Graduate University for Advanced Studies, SOKENDAI,
Tsukuba 305-0801, Japan\\
Yukawa Institute for Theoretical Physics, Kyoto University,
Kyoto 606-8502, Japan \\
}

\end{center}

\vspace*{1.0cm}

\begin{abstract}
{\normalsize \noindent
This work investigates the use of linear Paul traps as quantum sensors
for detecting megahertz gravitational waves.
Single-ion configurations exploit graviton-photon conversion in the
presence of external magnetic fields, while two-ion systems use
relative-motion excitations, which do not require magnets, to distinguish
gravitational waves from axion dark matter.
Furthermore, we show that entanglement of $N$ vibrational qubits enhances
the signal probability by a factor of $N^2$, improving sensitivity
beyond the standard quantum limit.
}
\end{abstract} 

%%%%%%%%%%%%%%%%%%%%%%%%%%%%%%%%%%%%%%%%%%%%%%%%%%%%%%%%%%%%%%%%%%%%%%%%%%%%
%%%%%%%%%%%%%%%%%%%%%%%%%%%%%%%%%%%%%%%%%%%%%%%%%%%%%%%%%%%%%%%%%%%%%%%%%%%%

\newpage
\baselineskip=18pt
\setcounter{page}{2}
\pagestyle{plain}

\setcounter{footnote}{0}

\tableofcontents
\noindent\hrulefill

%%%%%%%%%%%%%%%%%%%%%%%%%%%%%%%%%%%%%%%%%%%%%%%%%%%%%%%%%%%%%%%%%%%%%%%%%%%%
%%%%%%%%%%%%%%%%%%%%%%%%%%%%%%%%%%%%%%%%%%%%%%%%%%%%%%%%%%%%%%%%%%%%%%%%%%%%

\section{Introduction}

High-frequency gravitational waves are compelling probes of the early
universe, which remain inaccessible to electromagnetic observations.
Such signals may reveal imprints of phenomena such as phase transitions,
cosmic strings, and preheating after inflation~\cite{Caprini:2018mtu}.
While stochastic gravitational wave backgrounds at megahertz frequencies
are strongly constrained by Big Bang nucleosynthesis
bound~\cite{Yeh:2022heq}, non-stochastic gravitational waves,
such as those potentially generated by light primordial black hole
mergers~\cite{Franciolini:2022htd}, exotic compact
objects~\cite{Giudice:2016zpa}, and black hole
superradiance~\cite{Brito:2015oca}, may also
exist\footnote{They are typically expected to produce strain amplitudes in the range
$\sim 10^{-20}$--$10^{-30}$,
depending on the source parameters and distance.}.
Several proposals have been made for detecting megahertz gravitational
waves~\cite{Goryachev:2014yra,chou2017mhz,Aggarwal:2020umq,Domcke:2024mfu,
Patra:2024eke}, but a large region of parameter space remains unexplored.
The detection of high-frequency gravitational waves requires the
development of entirely new experimental
strategies~\cite{Aggarwal:2025noe}.

Quantum technologies have advanced rapidly across diverse architectures,
including superconducting circuits, trapped atoms, and photonic platforms,
enabling precise control of quantum states.
These developments are expected to have a profound impact in various
domains.
For example, quantum computers can solve certain problems more efficiently
than classical computers by exploiting superposition
and entanglement of qubits, the quantum analogs of classical
bits~\cite{ladd2010quantum}.
Another particularly promising direction for high-energy physics is
quantum sensing.
Quantum techniques can surpass the precision and resolution of classical
approaches, providing unprecedented sensitivity to weak signals such as
wave-like dark matter and high-frequency gravitational
waves~\cite{Degen:2016pxo}.

In this work, we focus on the use of linear Paul traps for the detection
of gravitational waves in the megahertz range, extending the framework
previously applied to searches for axion and dark photon dark
matter~\cite{Ito:2023zhp}.
A natural extension of these ideas involves graviton-photon conversion,
in which a gravitational wave is converted into an electric field resonant
with the center-of-mass oscillation of trapped ions.
Considering a trap with two ions, the relative motion is useful even in
the absence of magnetic fields, as gravitational waves periodically
expand and contract the ion separation, while axion dark matter would not
induce such an effect.
Furthermore, excitations induced by gravitational waves can surpass the
standard quantum limit when maximal entanglement is employed.

%%%%%%%%%%%%%%%%%%%%%%%%%%%%%%%%%%%%%%%%%%%%%%%%%%%%%%%%%%%%%%%%%%%%%%%%%%%%
%%%%%%%%%%%%%%%%%%%%%%%%%%%%%%%%%%%%%%%%%%%%%%%%%%%%%%%%%%%%%%%%%%%%%%%%%%%%

\section{Linear Paul traps}
\label{sec:iontrap}

The linear Paul trap is one of the most promising platforms for both
quantum computation and quantum sensing~\cite{bruzewicz2019trapped}.
Ions are confined in an ultrahigh-vacuum environment by a combination of
static and radio-frequency electric fields, forming a chain of coupled
harmonic oscillators aligned along the axis of the alternating-current
voltages, conventionally referred to as the $z$ axis.
Singly ionized atoms such as beryllium, calcium, barium, and ytterbium
are widely employed owing to their favorable ionization and
level-structure properties.
The two-level electronic states of these ions, arising from optical,
fine-structure, hyperfine, or Zeeman splittings, form the basis of
computational qubits.
We collectively refer to these two-level systems as spin qubits, with
$\Ket{g}$ and $\Ket{e}$ denoting the ground and excited states,
respectively.

A distinctive feature of Paul traps is their use of quantized vibrational
motion of trapped ions to mediate control of internal
states~\cite{Cirac:1995zz}.
The ions arrange themselves at intervals of the order of $1$--$10$~µm,
determined by the balance between Coulomb repulsion and external
trapping fields.
After laser cooling, the system exhibits quantized motional degrees of
freedom, whose collective oscillations are described as quantum harmonic
modes.
In what follows, we focus on oscillations along the $z$ direction.

To illustrate the mechanism, we first review the single-ion case.
We restrict attention to the lowest two motional states: the ground state
($\Ket{n = 0}$) and the first excited state ($\Ket{n = 1}$).
These states constitute what we call a vibrational qubit.
The Hamiltonian is
\begin{equation}
    H_0 = \frac{\omega_0}{2} \, \hat{\sigma}_z + \omega_z \,
    \hat{a}^\dagger \hat{a},
\end{equation}
where $\omega_z$ ($\sim$ MHz) is the vibrational frequency, $\omega_0$
($\sim$ GHz) is the energy splitting between $\Ket{g}$ and $\Ket{e}$, and
$\hat{\sigma}_z$ is the third Pauli operator acting on the spin qubit.
Here, $\hat{a}^\dagger$ and $\hat{a}$ denote creation and annihilation
operators of the vibrational mode, while $\hat{\sigma}_\pm$ act as spin
raising and lowering operators (e.g., $\hat{a}^\dagger \hat{\sigma}_-
\Ket{e,0} = \Ket{g,1}$).

Lasers tuned to specific frequencies allow precise control of ion states.
In the interaction picture, the effective Hamiltonian describing the
coupling between the ion and a laser of frequency $\omega$ and phase
$\phi$ is~\cite{walls2008quantum}
\begin{equation}
    H_{\rm laser} = \frac{\Omega}{2} \hat{\sigma}_+ e^{-i (\omega -
    \omega_0) t + i \phi} + \frac{i}{2} \eta \Omega \, \hat{a}^\dagger
    \left( \hat{\sigma}_+ e^{-i (\omega - \omega_0 - \omega_z) t + i
    \phi} - \hat{\sigma}_- e^{i (\omega - \omega_0 + \omega_z) t - i
    \phi} \right) + {\rm h.c.},
\end{equation}
where $\Omega$ is the Rabi frequency and $\eta$ is the Lamb--Dicke
parameter, typically small ($\sim 0.01$--$0.1$), which justifies
truncating higher-order terms.
Depending on the laser frequency, three resonances can be selectively
addressed: the red sideband transition ($\Ket{g, 1} \leftrightarrow
\Ket{e, 0}$) at $\omega = \omega_0 - \omega_z$, the blue sideband
transition ($\Ket{g, 0} \leftrightarrow \Ket{e, 1}$) at $\omega =
\omega_0 + \omega_z$, and the carrier transition ($\Ket{g, 0}
\leftrightarrow \Ket{e, 0}$) at $\omega = \omega_0$.
Together, these three resonances enable arbitrary single-qubit gates for
quantum computation, such as the Hadamard gate, $\Ket{g} \to (\Ket{g} +
\Ket{e}) / \sqrt{2}$, $\Ket{e} \to (\Ket{g} - \Ket{e}) / \sqrt{2}$.

With two ions in the trap, motion along the $z$ axis decomposes into
center-of-mass and stretch modes.
The center-of-mass mode is a quantum harmonic oscillation with effective
mass $2 m_{\rm ion}$ and frequency $\omega_{\rm cm} = \omega_z$, allowing
information transfer between ions and enabling two-qubit gates such as the
controlled-NOT (CNOT) gate: $\Ket{g, g} \to \Ket{g, g}$, $\Ket{g, e} \to
\Ket{g, e}$, $\Ket{e, g} \to \Ket{e, e}$, and $\Ket{e, e} \to \Ket{e, g}$.
The equilibrium separation between the two ions is given by
\begin{equation}
    d \equiv \left( \frac{2 \alpha_{\rm EM}}{m_{\rm ion}
    \omega_{\rm cm}^2} \right)^{1/3},
\end{equation}
where $\alpha_{\rm EM} = e^2 / 4 \pi$ is the fine-structure constant.
For reference, $d = (16, 3.5, 0.74)$~µm at $\omega_{\rm cm} = 2 \pi
\times (0.1, 1, 10)$~MHz with $^{171}$Yb$^+$ ions.
The small deviation around this stationary point, $z$, can be described as
the harmonic oscillation with effective mass $m_{\rm ion}/2$ and frequency
$\omega_{\rm rel} = \sqrt{3} \, \omega_{\rm cm}$, neglecting $\mathcal{O}
(z^3)$ terms.
In Secs.~\ref{sec:pair} and \ref{sec:enhance}, we denote the
relative-motion states $\Ket{n = 0}$ and $\Ket{n = 1}$ as a vibrational
qubit, again neglecting higher motional levels.

%%%%%%%%%%%%%%%%%%%%%%%%%%%%%%%%%%%%%%%%%%%%%%%%%%%%%%%%%%%%%%%%%%%%%%%%%%%%
%%%%%%%%%%%%%%%%%%%%%%%%%%%%%%%%%%%%%%%%%%%%%%%%%%%%%%%%%%%%%%%%%%%%%%%%%%%%

\section{Signals and sensitivities}

In this section, we investigate detection schemes for gravitational waves
and estimate their corresponding sensitivities.
First, we utilize the center-of-mass mode to probe effective electric
fields induced by gravitational waves via graviton-photon conversion,
following the approach previously employed for axion-like
particles~\cite{Ito:2023zhp}.
Next, we exploit the relative motion of the two-ion system, which enables
detection without requiring a strong external magnetic field, in contrast
to the center-of-mass method.
Finally, we show that entanglement of vibrational qubits can be used to
enhance the signal beyond the standard quantum limit.

%%%%%%%%%%%%%%%%%%%%%%%%%%%%%%%%%%%%%%%%%%%%%%%%%%%%%%%%%%%%%%%%%%%%%%%%%%%%

\subsection{Graviton-photon conversion and the center-of-mass mode}
\label{sec:cm}

In our analysis, we treat the gravitational wave as a weak perturbation
around the background Minkowski metric and expand relevant quantities
in the metric fluctuation $h_{\mu\nu}$ up to linear order.
Under the transverse-traceless gauge, the metric takes the form
\begin{equation}
    h^{\rm TT}_{ij}({\bm x}, t) = h^+ e^+_{ij} \cos (\omega t
    - {\bm k} \cdot {\bm x} + \phi^+) + h^\times e^\times_{ij}
    \cos (\omega t - {\bm k} \cdot {\bm x} + \phi^\times),
\end{equation}
where $\omega$ is the angular frequency, ${\bm k} = (\omega \sin\theta,
0, \omega \cos\theta)$ is the wave vector, and
\begin{equation}
    e^+_{ij} = \begin{pmatrix}
        \cos^2 \theta & 0 & -\cos \theta \sin \theta \\ 0 & -1 & 0
        \\ -\cos \theta \sin \theta & 0 & \sin^2 \theta
    \end{pmatrix}, \quad
    e^\times_{ij} = \begin{pmatrix}
        0 & \cos \theta & 0 \\
        \cos \theta & 0 & -\sin \theta \\ 0 & -\sin \theta & 0
    \end{pmatrix}
\end{equation}
represent the polarization tensors.
We consider monochromatic gravitational waves for simplicity.

Graviton-photon conversion is a phenomenon in which gravitational waves
convert into photons in the presence of strong external magnetic fields.
Under a electromagnetic background $\bar{A}^\mu$, the effective
Lagrangian describing the interaction between the graviton field
$h_{\mu\nu}$ and the induced electromagnetic field $A^\mu$ is
$\mathcal{L}_{\rm int} = j^\mu_{\rm eff} A_\mu$, where
\begin{equation}
    j^\mu_{\rm eff} = \partial_\nu \left( h^\mu_{\,\lambda}
    \bar{F}^{\lambda\nu} + h^\nu_{\,\lambda} \bar{F}^{\mu\lambda} \right)
    - \frac{1}{2} (\partial_\lambda h^\nu_{\,\nu}) \bar{F}^{\mu\lambda}
\end{equation}
is the effective current with the field strength tensor
$\bar{F}^{\mu\nu} = \partial^\mu \bar{A}^\nu - \partial^\nu
\bar{A}^\mu$~\cite{Ratzinger:2024spd}.
To describe the observable electromagnetic fields in the detector, we
adopt the proper detector frame, i.e., Fermi normal coordinates
constructed along the detector’s worldline, which provide a natural
description in the long-wavelength regime.

We focus on a static and uniform magnetic field ${\bm B} = (0, 0, B_z)$
inside a cylinder of radius $R$ and height $L$.
To define the quantization axis of spins, a weak magnetic field
$(\sim 1~{\rm mT})$ along $z$ axis are used in real Paul
traps~\cite{Pogorelov:2021eha}.
We expect that the region in which the external magnetic field is
applied can be extended independently of the Paul trap system by
employing, for example, superconducting magnets.
In this case, the parameter $R$ is typically on the order of meters.
The wavelength of a megahertz gravitational wave is about 300~m, much
longer than the experimental length scale, and thereby the proper detector
frame is well suited.
At the leading order of $\omega R$, the induced electric field along $z$
axis is
\begin{equation}
    E_z = - h^\times B_z (\omega R)^2 \, f (L / 2 R) \,
    \sin^2 \theta \cos (\omega t + \phi^\times) ,
    \label{eq:gpconv}
\end{equation}
where the logarithmic factor $f (x) = \log (x + \sqrt{1 + x^2})$,
determined by the geometry of the magnet, contributes an
$\mathcal{O}(1)$ correction around $x \sim 1$.

The interaction Hamiltonian of the single vibrational qubit with the
electric field induced by a gravitational wave is expressed as
\begin{equation}
    H_{\rm int} = \frac{e E_z}{\sqrt{2 m_{\rm ion} \omega_z}} \left(
    \hat{a}^\dagger e^{i \omega_z t} + \hat{a} e^{-i \omega_z t} \right),
\end{equation}
which becomes resonant when $\omega = \omega_z$.
Then, using the rotating-wave approximation, the time evolution of the
vibrational qubit is governed by the displacement operator $D (\beta)
= \exp (\beta \, \hat{a}^\dagger - \beta^* \hat{a})$ with
\begin{equation}
    \beta = \alpha T, \quad \alpha = \frac{h^\times e^{i \phi}}{2}
    \frac{e B_z}{\sqrt{2 m_{\rm ion} \omega_z}} (\omega_z R)^2
    \, f (L / 2 R) \, \sin^2 \theta.
\end{equation}
Here, $T$ is the observation time, and $\phi$ absorbs all complex phases.

Starting from the initial state $\Ket{{\rm init}} = \Ket{g, 0}$, the
action of this operator evolves the vibrational state into
\begin{equation}
    D (\beta) \Ket{g, 0} = \cos \vert \beta \vert \, \Ket{g, 0} +
    e^{i \phi} \sin \vert \beta \vert \Ket{g, 1} \simeq \Ket{g, 0}
    + \beta \Ket{g, 1},
    \label{eq:timeevolution}
\end{equation}
yielding an excitation probability of $\vert \beta \vert^2 = \vert
\alpha \vert^2 T^2$.
The vibrational state is then mapped onto the spin state by applying
a red-sideband $\pi$-pulse, $\Ket{g, 1} \to \Ket{e, 0}$.
Subsequently, the spin state is measured using standard protocols, for
example, by observing fluorescence after applying a laser that excites
one of the spin states into a short-lived auxiliary
level~\cite{Bernardini:2023yir}.
The measurement is performed after an interrogation time $T$ and repeated
$N_{\rm shot}$ times, such that the total duration of the experiment
is approximately $T_{\rm total} \simeq N_{\rm shot} T$.
The expected number of signal events induced by gravitational waves is
therefore $S = N_{\rm shot} \vert \alpha \vert^2 T^2$.

In actual configurations, the vibrational qubit is perturbed primarily
by thermal photons originating from the trap
electrodes~\cite{brownnutt2015ion}.
The contribution of the noise increases linearly over $T$, characterized
by the heating rate $\dot{\bar{n}}$.
The number of noise events is given by $B = N_{\rm shot} \dot{\bar{n}} T$,
as discussed in Ref.~\cite{Ito:2023zhp}.
Defining the sensitivity through the criterion $S / \sqrt{B} > 1.645$
(corresponding to the 95\% confidence level), we obtain the sensitivity
to the amplitude of electric field with angular frequency $\omega =
\omega_z$ as
\begin{equation}
    E_0 = 1.5~{\rm nV/m} \times
    \left( \frac{\dot{\bar{n}}}{0.1~{\rm Hz}} \right)^{\! 1/4}
    \left( \frac{T_{\rm total}}{1~{\rm day}} \right)^{\! -1/4}
    \left( \frac{T}{1~{\rm s}} \right)^{\! -1/2}
    \left( \frac{m_{\rm ion}}{37~{\rm GeV}} \right)^{\! 1/2}
    \left( \frac{\omega_z / 2 \pi}{1~{\rm MHz}} \right)^{\! 1/2},
\end{equation}
where we assume the use of a $^{40}$Ca$^+$ ion.
The reference heating rate of 0.1~Hz is several times lower than
the best values reported in Ref.~\cite{brownnutt2015ion}.
Such a low rate is necessary to satisfy the condition $\dot{\bar{n}} T
\ll 1$ for interrogation times of order one second.
In practice, the heating rate decreases with larger trap sizes and lower
temperatures.
Therefore, both the trap size and cryogenic operation should be optimized
to maximize sensitivity.

This detection scheme can thus be employed to probe gravitational waves,
which induce weak electric fields as described in Eq.~\eqref{eq:gpconv}.
The corresponding strain sensitivity for a single-ion system is
obtained as
\begin{equation}
    \begin{split}
        h^\times = 1.7 &\times 10^{-12} \times \left(
        \frac{\dot{\bar{n}}}{0.1~{\rm Hz}} \right)^{\! 1/4}
        \left( \frac{T_{\rm total}}{1~{\rm day}} \right)^{\! -1/4}
        \left( \frac{T}{1~{\rm s}} \right)^{\! -1/2} \\
        &\times \left( \frac{m_{\rm ion}}{37~{\rm GeV}} \right)^{\! 1/2}
        \left( \frac{\omega_z / 2 \pi}{1~{\rm MHz}} \right)^{\! -3/2}
        \left( \frac{B_z}{1~{\rm mT}} \right)^{\! -1} \left(
        \frac{R}{3~{\rm m}} \right)^{\! -2} \left(
        \frac{f (L / 2 R)}{1} \right)^{\! -1}
    \end{split}
    \label{eq:gpconv_sensitivity}
\end{equation}
at the 95\% confidence level.
In deriving Eq.~\eqref{eq:gpconv_sensitivity}, we have averaged over the
polarization angle $\theta$.
Furthermore, by tuning the resonance frequency of the vibrational mode,
one can scan over a broad frequency range of gravitational waves.
In practice, the resonance frequency may be tunable within the range
100~kHz to 10~MHz.

Paul traps also provide a versatile platform for generating entanglement
among qubits, making them suitable for quantum computing and
quantum-enhanced sensing.
As shown in Ref.~\cite{Ito:2023zhp}, a maximally entangled state of $N$
vibrational qubits, prepared through a sequence of quantum operations,
such as the red-sideband interaction, the Hadamard and CNOT gates, in a
Paul trap, enhances the excitation rate by a factor of $N^2$ compared
to the single-ion case.

There are two principal approaches to preparing multiple vibrational
qubits and enabling their interaction.
The first involves transporting ions and performing gate operations on
individual vibrational qubits after generating entanglement among spin
qubits.
In this case, one must carefully account for gate fidelities and the
heating rate in an $N$-ion detector, which generally degrade relative to
the single-ion case~\cite{bruzewicz2019trapped}.
The second approach employs a network of interconnected Paul traps, each
containing a single ion~\cite{Kielpinski:2002wbd}.
In this article, we assume that the latter technology is available.
Under this assumption, thermal noise photons act
incoherently on different ions, and the strain sensitivity is expected to
improve as $N^{3/4}$.

%%%%%%%%%%%%%%%%%%%%%%%%%%%%%%%%%%%%%%%%%%%%%%%%%%%%%%%%%%%%%%%%%%%%%%%%%%%%

\subsection{The stretch mode of a pair of ions}
\label{sec:pair}

In this and the following subsections, we consider two-ion detectors.
The center-of-mass mode provides a feasible means of searching for gravitational waves.
However, the resulting signal is indistinguishable from that produced
by axion dark matter~\cite{Ito:2023zhp}.
To discriminate between the two cases, an alternative method for
gravitational wave detection is required.
As discussed in Sec.~\ref{sec:iontrap}, the relative motion of the ions
offers such a possibility.
After laser cooling~\cite{lechner2016electromagnetically}, this mode is
quantized and described as a quantum simple harmonic oscillator with an
effective mass of $m_{\rm ion} / 2$ and an angular frequency
$\omega_{\rm rel} = \sqrt{3} \, \omega_{\rm cm}$.
Importantly, this strategy does not require the presence of external
magnetic fields.

The non-relativistic Hamiltonian, expanded up to order $m_{\rm ion}$,
is~\cite{Ito:2020xvp}\footnote{Here we work in the proper
detector frame, where the leading coupling between the relative
displacement of two ions and spacetime curvature is described by the
Riemann tensor. It remains valid for gravitational waves in the
long-wavelength limit, where the wavelength is much larger than the ion
separation.}
\begin{equation}
    H = \frac{1}{2} m_{\rm ion} R_{0k0l} \, x^k x^l
    \supset \frac{h^+}{4} m_{\rm ion} \omega^2 \sin^2 \theta
    \, e^{i (\omega t + \phi^+)} \, d \, z + {\rm h.c.},
\end{equation}
with $R_{\mu\nu\rho\sigma}$ being the Riemann tensor evaluated at the
origin ${\bm x} = 0$, corresponding to the center of mass of the ions.
The indices $k$ and $l$ run over spatial directions, and $z$ denotes the
displacement of the distance between two ions.
Terms of order lower than $m_{\rm ion}$ are neglected.
A resonance arises when $\omega = \omega_{\rm rel}$, yielding the
effective Hamiltonian
\begin{equation}
    H_{\rm int} = \left[ \frac{h^+}{2} e^{i \phi} \left( \frac{9}{16}
    \alpha_{\rm EM}^2 m_{\rm ion} \omega_{\rm rel}^5 \right)^{1/6}
    \sin^2 \theta \right] \hat{a}^\dagger + {\rm h.c.} \equiv \epsilon
    \hat{a}^\dagger + {\rm h.c.}
\end{equation}
under the rotating-wave approximation.
In this expression, $\hat{a}^{(\dagger)}$ are the ladder operators of the
stretch mode, which we refer to as the vibrational qubit henceforth.
A real parameter $\phi$, encapsulating all complex phases, is assumed to
be randomly distributed.
The time evolution is then described by the displacement operator
$D (\beta) = \exp (\beta \, \hat{a}^\dagger - \beta^* \hat{a})$ with
$\beta = -i \epsilon T$.

The signal corresponds to an excitation of the vibrational mode from
the ground state $\Ket{n = 0}$ to the first excited state
$\Ket{n = 1}$, as described in Eq.~\eqref{eq:timeevolution}.
This vibrational excitation is mapped onto the spin state by applying
a red-sideband $\pi$-pulse.
The spin state is subsequently measured using standard detection
protocols, for instance, fluorescence observation after driving one of
the spin levels to a short-lived auxiliary state with a laser.
The probability of observing the signal is given by $\vert \epsilon
\vert^2 T^2$, where $T$ denotes the observation time per single shot.
As in the previous subsection, the measurement is repeated
$N_{\rm shot} = T_{\rm total} / T$ times.
The number of signal events is therefore $S = N_{\rm shot} \vert \epsilon
\vert^2 T^2$, while the number of the noise events is $B = N_{\rm shot}
\dot{\bar{n}} T$, with $\dot{\bar{n}}$ being the heating rate of the
stretch mode.
It has been reported that the heating rate of the stretch mode is lower
than that of the center-of-mass mode in other Paul trap
architectures~\cite{King:1998si,lechner2016electromagnetically}.
The sensitivity, defined by the condition $S / \sqrt{B} > 1.645$
corresponding to the 95\% confidence level, is obtained as
\begin{equation}
    \begin{split}
        h^+ = 3.6 \times 10^{-11} \times
        \left( \frac{\dot{\bar{n}}}{0.1~{\rm Hz}} \right)^{\! 1/4}
        \left( \frac{T_{\rm total}}{1~{\rm day}} \right)^{\! -1/4}
        \left( \frac{T}{1~{\rm s}} \right)^{\! -1/2} \left(
        \frac{m_{\rm ion}}{159~{\rm GeV}} \right)^{\! -1/6} \left(
        \frac{\omega_{\rm cm} / 2 \pi}{1~{\rm MHz}} \right)^{\! -5/6} ,
    \end{split}
\end{equation}
where we consider $^{171}$Yb$^+$ ions and average over the polarization
angle $\theta$.
Notably, heavier ions are advantageous for this method, in contrast to
detection schemes based on graviton-photon conversion.

%%%%%%%%%%%%%%%%%%%%%%%%%%%%%%%%%%%%%%%%%%%%%%%%%%%%%%%%%%%%%%%%%%%%%%%%%%%%

\subsection{Quantum enhancement with multiple ion pairs}
\label{sec:enhance}

In this subsection, we demonstrate that a maximally entangled state of $N$
vibrational qubits, generated through a sequence of quantum operations in
Paul traps, enhances the excitation rate by a factor of $N^2$ relative to
the single-ion case.
Assuming the availability of technology that permits state manipulation
of qubits distributed across separate traps, we consider a configuration
in which $N$ qubits, corresponding to $2 N$ ions, are arranged in $N$
two-ion detectors.
We label the ions in the $j$th two-ion detector as A$_j$ and B$_j$.
The state of the $j$th detector is characterized by three qubits: the spin
qubits of A$_j$ and B$_j$, and the vibrational qubit.

The time evolution of the ions induced by gravitational waves, which
resonantly couple to the vibrational qubits, is described by the
displacement operator $D (\beta) = \prod_j D_j (\beta)$, $D_j (\beta) =
\exp (\beta \, \hat{a}^\dagger_j - \beta^* \hat{a}_j)$, where
$\hat{a}_j^{(\dagger)}$ are the ladder operators acting on the $j$th
vibrational qubit.
The coupling $\beta = \beta_{\rm r} + i \beta_{\rm i}$ is assumed to be
sufficiently small so that quadratic terms can be neglected.
In this limit, the operator can be rewritten as
\begin{equation}
    D_j (\beta) = \left( 1 + i \beta_{\rm i} \hat{\sigma}_j^1 \right)
    e^{-i \beta_{\rm r} \hat{\sigma}_j^2} = \left( 1 - i \beta_{\rm r}
    \hat{\sigma}_j^2 \right) e^{i \beta_{\rm i} \hat{\sigma}_j^1},
\end{equation}
with $\hat{\sigma}_j^1 = \hat{a}_j^\dagger + \hat{a}_j$ and
$\hat{\sigma}_j^2 = i \hat{a}_j^\dagger - i \hat{a}_j$ representing the
first and second Pauli matrices acting on the $j$th vibrational qubit.

We describe a protocol for measuring the imaginary part $\beta_{\rm i}$.
The procedure begins with the ground state $\Ket{{\rm init}} \equiv
\Ket{g, g, 0}^{\otimes N}$.
A Hadamard gate is first applied to ion B$_1$, followed by CNOT gates
on every pair $({\rm B}_1, {\rm B}_j)$, $j = 1, 2, \dots, N$.
This sequence prepares the state
\begin{equation}
    \frac{1}{\sqrt{2}} \left[ \Ket{g, g, 0}^{\otimes N}
    + \Ket{g, e, 0}^{\otimes N} \right] .
\end{equation}
At this stage, the B ions are maximally entangled in the state commonly
referred to as the Greenberger--Horne--Zeilinger state.
By subsequently applying Hadamard gates to the B ions and the
red-sideband operators $(\Ket{e, 0} \to \Ket{g, 1})$ on all ion pairs,
the system is transformed into
\begin{equation}
    \Ket{\psi (0)} = \frac{1}{\sqrt{2}} \left[ \Ket{+}^{\otimes N}
    + \Ket{-}^{\otimes N} \right] ,
\end{equation}
where $\Ket{\pm} = (\Ket{g, g, 0} \pm \Ket{g, g, 1}) / \sqrt{2}$ are
eigenstates of $\hat{\sigma}_j^1$ with eigenvalues $\pm 1$.

The gravitational wave acts on the ions, yielding
\begin{equation}
    \Ket{\psi (T)} = D (\beta) \Ket{\psi (0)} = \frac{1}{\sqrt{2}}
    \left[ e^{i N \beta_{\rm i}} \big( \Ket{+} - \beta_{\rm r} \Ket{-}
    \big)^{\otimes N} + e^{-i N \beta_{\rm i}} \big( \Ket{-} +
    \beta_{\rm r} \Ket{+} \big)^{\otimes N} \right]
\end{equation}
with $\beta = - i \epsilon T$.
By applying the inverse of the state-preparation operators, we obtain
\begin{equation}
    \Ket{{\rm fin}} = \big( \Ket{g, g, 0} + i N \beta_{\rm i}
    \Ket{g, e, 0} \big) \otimes \Ket{g, g, 0}^{\otimes N-1} + \cdots,
\end{equation}
where the omitted terms are orthogonal to the first term.
Since these orthogonal components do not contribute to the measurement
of the target state, they can be neglected in the following analysis.
The sequence of operations returns the system to its initial state,
$\Ket{{\rm fin}} = \Ket{{\rm init}}$, if $\beta = 0$.

The signal event is the excitation of the B$_1$ spin state, and its
probability is enhanced by a factor of $N^2$:
\begin{equation*}
    \left\vert \left( \Bra{g, e, 0} \otimes \Bra{g, g, 0}^{\otimes N-1}
    \right) \Ket{{\rm fin}} \right\vert^2 = N^2 \beta_{\rm i}^2.
\end{equation*}
This enhancement arises from quantum interference in the maximally
entangled state, where the collective contribution of $N$ qubits
coherently amplifies the excitation probability compared to the
single-ion case.
To measure the the real part $\beta_{\rm r}$, one should prepare the state
\begin{equation}
    \Ket{\psi' (0)} = \frac{1}{\sqrt{2}} \left[ \left( \frac{\Ket{g, g, 0}
    + i \Ket{g, g, 1}}{\sqrt{2}} \right)^{\otimes N} + \left( \frac{\Ket{g,
    g, 0} - i \Ket{g, g, 1}}{\sqrt{2}} \right)^{\otimes N} \right] ,
\end{equation}
composed of the eigenstates of $\hat{\sigma}^2_j$, instead of
$\Ket{\psi (0)}$.
In both measurements, the procedures are repeated $N_{\rm shot}$ times.
Since the phase of gravitational wave is unknown, we take an average:
$\overline{\beta^2_{\rm r}} = \overline{\beta_{\rm i}^2} = \vert
\epsilon \vert^2 T^2 / 2$.

If heating noise excites the ion oscillations incoherently, the
excitation rate scales linearly with $N$.
Consequently, the significance of the signal scales as
\begin{equation}
    \frac{S}{\sqrt{B}} \propto N^{3/2}.
\end{equation}
The sensitivity is improved accordingly, yielding $h^+ = 5.1 \times
10^{-11} \times N^{-3/4}$ under the same parameters as in
Sec.~\ref{sec:pair}.
We implicitly assume that the system coherence time,
including the lifetimes of the excited states and the decoherence time of
the entanglement, is much longer than the single-shot observation time\footnote{
A coherence time of about 50~s has been demonstrated
in a trapped-ion experiment~\cite{Harty:2014tkj}.}.
While the entanglement-enhanced scheme is experimentally challenging with
current technology~\cite{Pogorelov:2021eha,Moses:2023ozv}, ongoing progress
in trapped-ion coherence times, gate fidelities, and large-scale
entanglement suggests that the required conditions could be explored
in future experiments.

%%%%%%%%%%%%%%%%%%%%%%%%%%%%%%%%%%%%%%%%%%%%%%%%%%%%%%%%%%%%%%%%%%%%%%%%%%%%
%%%%%%%%%%%%%%%%%%%%%%%%%%%%%%%%%%%%%%%%%%%%%%%%%%%%%%%%%%%%%%%%%%%%%%%%%%%%

\section{Summary}

We investigate the use of linear Paul traps as quantum sensors for
detecting high-frequency gravitational waves, which are probes of the
early universe.
A linear Paul trap confines ions in an ultrahigh vacuum using a
combination of static and radio-frequency electric fields, forming a
chain along the trap axis.
The quantized vibrational modes of ions and internal two-level electronic
states serve as vibrational and spin qubits, enabling precise quantum
control for computation and sensing applications.

Single-ion configurations exploit graviton-photon conversion in the
presence of strong external magnetic fields, where a gravitational wave
induces an effective electric field that couples to the vibrational
qubit of the ion.
The vibrational excitation is mapped onto the spin state with a laser,
and fluorescence measurements allow extraction of the signal.
Two-ion systems utilize not only the center-of-mass mode but also the
relative motion between ions, which can be affected by gravitational
waves but not by axion dark matter, allowing discrimination between
the two.
By tuning the resonance frequency of the vibrational mode, a range of
gravitational wave frequencies, from hundreds of kilohertz to several
megahertz, can be scanned.

The use of multiple entangled vibrational qubits across separate Paul
traps enables quantum enhancement of the signal.
By preparing maximally entangled states, the excitation probability of
the vibrational modes scales as $N^2$ for $N$ qubits, leading to a 
signal-to-noise improvement proportional to $N^{3/2}$.
This method allows the system to probe strains of gravitational waves
beyond the standard quantum limit.
It is required that the decoherence time and the lifetime of qubit
excitations exceed the observation time, and that high-fidelity
operations capable of connecting a large number of ions are anticipated
to become available in the near future.

%%%%%%%%%%%%%%%%%%%%%%%%%%%%%%%%%%%%%%%%%%%%%%%%%%%%%%%%%%%%%%%%%%%%%%%%%%%%
%%%%%%%%%%%%%%%%%%%%%%%%%%%%%%%%%%%%%%%%%%%%%%%%%%%%%%%%%%%%%%%%%%%%%%%%%%%%

\section*{Acknowledgment}

We would like to thank Asuka Ito, Ryuichiro Kitano and Wakutaka Nakano
for useful discussion.
The work is supported by JSPS KAKENHI Grant Number JP24KJ1157.

%%%%%%%%%%%%%%%%%%%%%%%%%%%%%%%%%%%%%%%%%%%%%%%%%%%%%%%%%%%%%%%%%%%%%%%%%%%%
%%%%%%%%%%%%%%%%%%%%%%%%%%%%%%%%%%%%%%%%%%%%%%%%%%%%%%%%%%%%%%%%%%%%%%%%%%%%

\bibliography{bibcollection}
\bibliographystyle{modifiedJHEP}

%%%%%%%%%%%%%%%%%%%%%%%%%%%%%%%%%%%%%%%%%%%%%%%%%%%%%%%%%%%%%%%%%%%%%%%%%%%%
%%%%%%%%%%%%%%%%%%%%%%%%%%%%%%%%%%%%%%%%%%%%%%%%%%%%%%%%%%%%%%%%%%%%%%%%%%%%

\end{document}